\documentclass[conference]{IEEEtran}
\IEEEoverridecommandlockouts

\usepackage{cite}
\usepackage{amsmath,amssymb,amsfonts}
\usepackage{graphicx}
\usepackage{textcomp}
\usepackage{xcolor}
\usepackage[T1]{fontenc}
\usepackage[utf8]{inputenc}

\usepackage[hidelinks,bookmarks=false]{hyperref}

\hypersetup{
  pdftitle={Mesa Orientation Engineering for Polarization Locking in VCSELs},
  pdfauthor={Zifeng Yuan, Wenkai Shan, Tingting Chen, Beibei Lin, Aaron Danner},
  pdfsubject={Vertical-cavity surface-emitting lasers; polarization switching; optical injection locking},
  pdfkeywords={mesa orientation, VCSEL, polarization switching, polarization locking, optical injection locking, Ising machine}
}

\def\BibTeX{{\rm B\kern-.05em{\sc i\kern-.025em b}\kern-.08em T\kern-.1667em\lower.7ex\hbox{E}\kern-.125emX}}

\begin{document}

\title{Mesa Orientation Engineering for Polarization\\ Locking in VCSELs}

\author{\IEEEauthorblockN{Zifeng Yuan, Wenkai Shan, Tingting Chen, Beibei Lin, and Aaron Danner}
\IEEEauthorblockA{\textit{Department of Electrical and Computer Engineering} \\
\textit{National University of Singapore}\\
4 Engineering Drive 3, 117583 Singapore \\
adanner@nus.edu.sg}
}

\maketitle

\begin{abstract}
While polarization stability is common in many optical systems, intentional polarization bi-stability is also important for polarization-encoded computing. We propose a mesa orientation strategy to engineer the polarization switching of vertical-cavity surface-emitting lasers (VCSELs). Further experiments demonstrate improved polarization locking performance.
\end{abstract}


\section{Introduction}

Vertical-cavity surface-emitting lasers (VCSELs) have emerged as widely used semiconductor lasers due to their compact footprint, low power consumption, and ease of integration into large-scale arrays. While polarization stability is common in optical systems, intentional polarization instability or bi-stability plays a critical role in emerging applications where polarization has been proposed as a means of encoding information, such as in Ising machines and optical neural networks~\cite{Utsunomiya2011, Babaeian2019}. In particular, recent studies have proposed using VCSELs to encode bits or qubits in orthogonal lasing polarization states~\cite{Loke2023, Yuan2025a}. Computation is then performed via optical injection locking (OIL)~\cite{Chang2003}, a technique long exploited to push the modulation bandwidth of semiconductor lasers well beyond the free-running limit~\cite{Zhao2007, Lau2008}, in which a master laser locks the polarization state of a slave VCSEL and thereby mediates controlled interactions between VCSEL-based qubits~\cite{Gao2024, Zhang2025}. Such systems have since been scaled into array-level prototypes~\cite{Lim2024, Zhang2025b} and dedicated device architectures~\cite{Zhang2025Patent}, paralleling integrated optical processors realized on other material platforms~\cite{Liu2025}.

The polarization response of a VCSEL under optical injection has been characterized in detail over the past two decades. Orthogonal injection can drive polarization switching accompanied by nonlinear dynamics and bistability~\cite{Gatare2006, Perez2011}, and injection-induced switching has been reported for both parallel and orthogonal injection at long wavelengths~\cite{Jeong2008, DenisleCoarer2017}. The extent of the locking region depends strongly on the injected state of polarization, with linear, elliptical and circular injection each producing distinct boundaries~\cite{Qader2011, AlSeyab2013, Lin2014}, while the polarization state is also sensitive to feedback conditions~\cite{Nazhan2017}. Beyond polarization control, OIL has been applied to long-wavelength VCSEL-by-VCSEL links~\cite{Hayat2009}, carrier recovery and comb generation~\cite{Jignesh2017, Prior2016}, spin-polarization modulation~\cite{Yokota2023}, and the coupling of emitters in large arrays~\cite{Pfluger2023}.

However, typically VCSELs inherently exhibit gain anisotropy caused by strain or crystal anisotropy, resulting in a polarization state with predominantly a higher gain~\cite{Gehrsitz2000}. This presents a challenge for polarization locking: in VCSELs with strong anisotropy, the polarization becomes fixed and difficult to switch, or achieving polarization locking requires significantly higher injection power. Controlling the cavity geometry is a long-established route to influencing this anisotropy: anisotropic transverse cavity geometries were shown early on to fix the polarization axis~\cite{Choquette1994b}, and anisotropic geometry combined with injection can induce switching outright~\cite{Tan2012}. Reducing the built-in anisotropy through the oxide aperture correspondingly lowers the injection power needed for locking~\cite{Yuan2025c}, and wafer-scale measurements confirm that such tailored apertures give reproducible polarization statistics across large device populations~\cite{Yuan2025b, Yuan2025e}, a prerequisite for the array-level operation these schemes require. The spin-flip model provides the standard theoretical framework for describing the resulting bistability between the two linearly polarized modes~\cite{Martin1997, AlSeyab2011}. As a result, large-scale photonic computing systems built from strongly anisotropic VCSELs may suffer from increased energy consumption, and co-integration with other photonic components imposes further constraints~\cite{Lim2025}.

In this work, we propose a mesa orientation strategy to control the polarization switching property of VCSELs. By aligning the elongated axis of the mesa either along the direction of inherent anisotropy or orthogonal to it, we achieve two distinct behaviors: stable $x$-polarization due to enhanced anisotropy, or polarization switching resulting from anisotropy compensation. Building on this observation, we further demonstrate improved polarization locking in the switchable VCSEL, which exhibits a broader locking range and requires lower injection power.

\section{Mesa Orientation Engineering}

For polarization-encoded applications, what we desire is a VCSEL without a built-in polarization preference---in other words, a VCSEL with reduced gain anisotropy compared to conventional ones. In the following, we shall show how this can be achieved using our proposed approach of rotating the mesa orientation.

\begin{figure}[!t]
\centering
\includegraphics[width=\columnwidth]{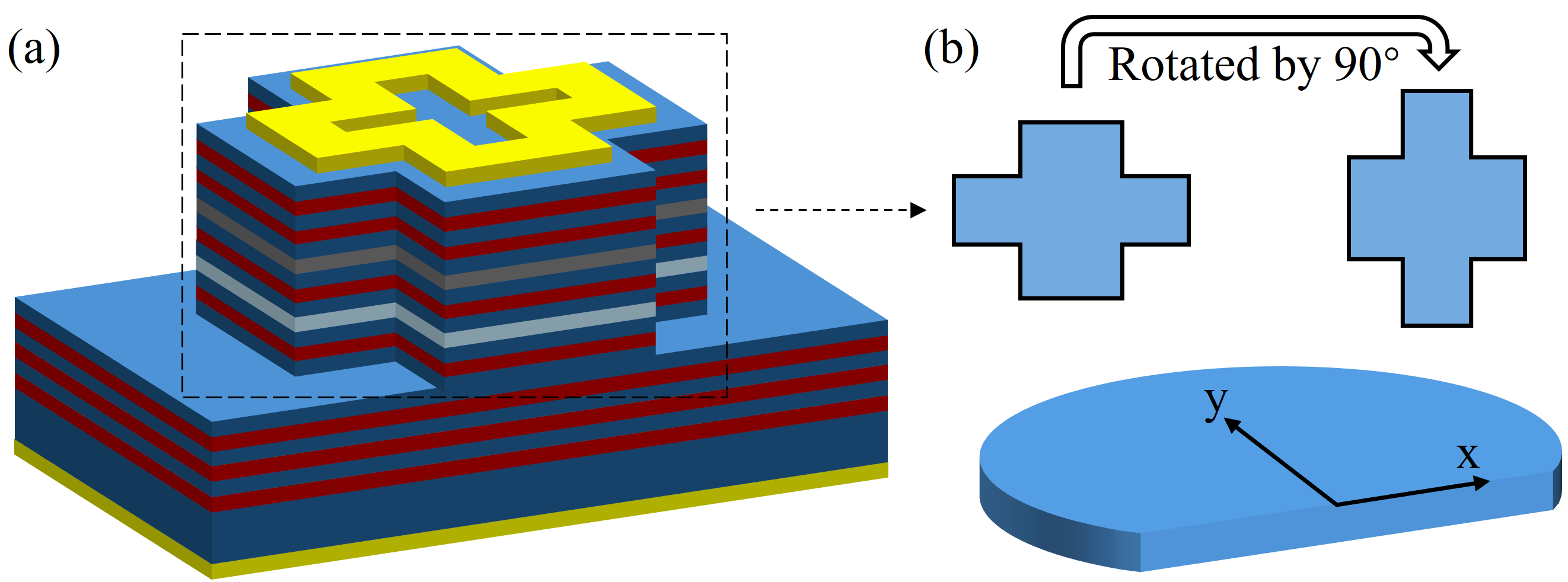}
\caption{(a) Schematic of a VCSEL with a cruciform mesa. (b) Mesa orientations with one aligned to the wafer flat ($x$-axis), and the other rotated by 90\textdegree{} to lie along the $y$-axis.}
\label{fig:fig1}
\end{figure}

The mesa is typically defined during the etching step of the fabrication process. As shown in Fig.~\ref{fig:fig1}, the black box in Fig.~\ref{fig:fig1}(a) highlights the etched mesa region within the VCSEL structure. In this work, the mesa is designed in a cruciform shape with an aspect ratio of 0.7 between the short and long arms. Our method involves fabricating two types of VCSELs: in one configuration, the long axis of the mesa is aligned parallel to the flat edge of the wafer (defined as the $x$-direction); in the other, it is rotated by 90\textdegree{}, making it orthogonal to the flat (i.e., along the $y$-direction). Following mesa etching, wet oxidation typically produces uniform oxidation rates in all directions. As a result, the final oxide aperture naturally mirrors the defined mesa geometry. By simply rotating the mesa orientation, we realize two different aperture geometries, whose polarization characteristics will be discussed in later measurements.

\begin{figure}[!t]
\centering
\includegraphics[width=\columnwidth]{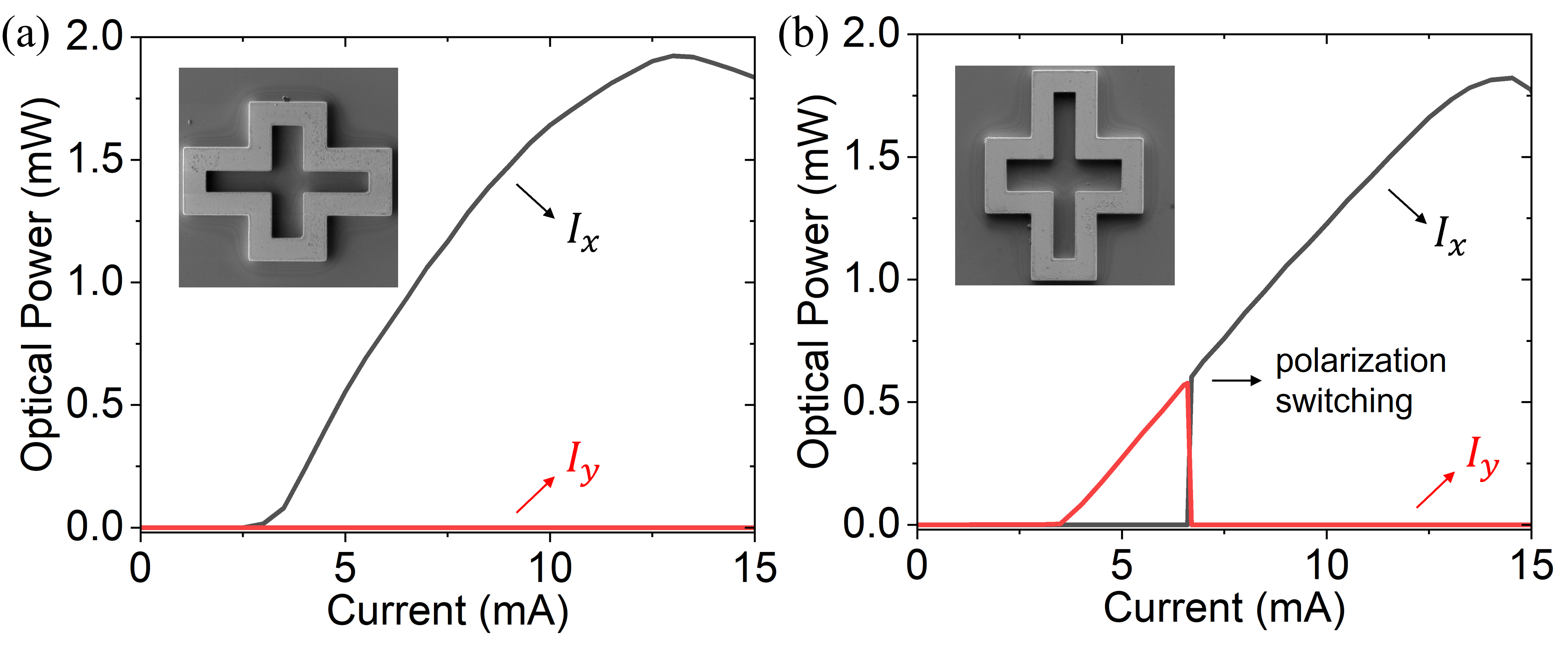}
\caption{Polarization-resolved light--current curves for VCSELs with (a) horizontal and (b) vertical elongated mesas.}
\label{fig:fig2}
\end{figure}

Next, we measure the polarization behavior of the two types of VCSELs. As shown in Fig.~\ref{fig:fig2}, the initial polarization states (just above threshold) of the two VCSELs are orthogonal and aligned with the direction of the elongated mesa arm. Upon increasing the injection current, the VCSEL with the 90\textdegree{} rotated mesa (elongated along the $y$-direction) exhibits polarization switching---from initial $y$ to $x$-polarization. This behavior arises from the inherent gain anisotropy in VCSELs, which typically favors $x$-polarized emission due to crystallographic or strain-induced asymmetries. By elongating the aperture along the $y$-direction, this anisotropy is compensated, resulting in more balanced gain between the $x$ and $y$-polarization states. Consequently, the polarization becomes less stable, and switching occurs under higher injection. Through this mesa-orientation approach, we are able to fabricate two types of VCSELs: one with a stable polarization state (Fig.~\ref{fig:fig2}(a)), and another that exhibits reduced anisotropy and polarization switching behavior (Fig.~\ref{fig:fig2}(b)).

\section{Improved Polarization Locking}

\begin{figure}[!t]
\centering
\includegraphics[width=\columnwidth]{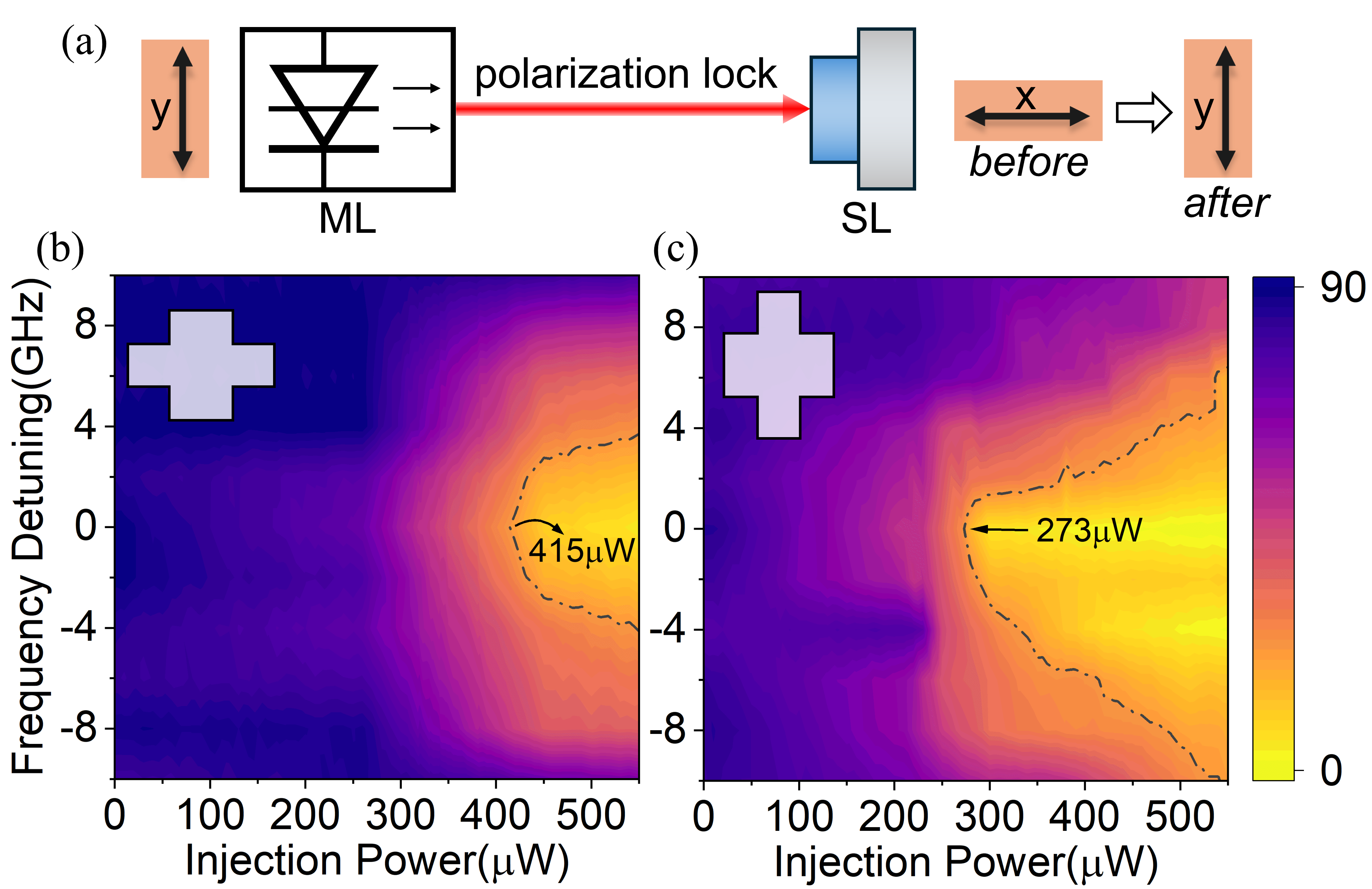}
\caption{(a) Polarization locking setup. (b, c) Locking maps for VCSELs with (b) horizontal and (c) vertical elongated mesas. Color indicates polarization azimuth difference between SL and ML ($\Delta\theta = \theta_{\mathrm{SL}} - \theta_{\mathrm{ML}}$).}
\label{fig:fig3}
\end{figure}

We further evaluate the polarization locking performance of the two VCSELs. The experimental setup is shown in Fig.~\ref{fig:fig3}(a). The fabricated VCSELs serve as slave lasers (SL), biased at a fixed current where they exhibit $x$-polarization. A master laser (ML) with linear $y$-polarization---orthogonal to that of the SL---is then injected into the cavity of our fabricated VCSELs. In this experiment, we vary both the frequency detuning ($\Delta f$ between ML and SL) and injection power of the ML while monitoring the polarization state of the SL. The results are shown in Figs.~\ref{fig:fig3}(b) and \ref{fig:fig3}(c), where the vertical axis represents the frequency detuning, and the horizontal axis denotes the injection power. The color bar denotes the difference in polarization azimuth between the SL and ML: $\Delta\theta = \theta_{\mathrm{SL}} - \theta_{\mathrm{ML}}$. A value of 0\textdegree{} indicates successful polarization locking (SL aligned with ML), while 90\textdegree{} indicates unsuccessful locking (SL remains orthogonal to ML).

Comparison of the two locking maps reveals that the VCSEL exhibiting polarization switching behavior (reduced anisotropy) in Fig.~\ref{fig:fig3}(c) achieves a broader locking range and requires lower injection power compared to the polarization-stable VCSEL in Fig.~\ref{fig:fig3}(b). This result underscores the benefit of anisotropy compensation for achieving efficient polarization locking.

\section{Conclusion}

In this work, we demonstrate a mesa orientation strategy to control the polarization switching behavior and enhance polarization locking in VCSELs. By aligning the mesa along two orthogonal directions, we achieve either stable polarization or polarization switching by enhancing or compensating the inherent gain anisotropy. Polarization locking experiments further confirm that VCSELs with vertically elongated mesas exhibit improved polarization locking performance with lower injection power. These findings highlight mesa orientation engineering as a practical approach for optimizing VCSEL-based polarization-encoded photonic systems. Reliable polarization encoding may further support emerging benchmarks and applications in computational imaging~\cite{lin2025rgb, Teng2025, Lin_2026_CVPR, lin2026geocomplete, cao20263dot}, low-level vision (e.g., image restoration and enhancement)~\cite{lin2024nightrain, lin2025nighthaze, chen2024dual, lin2025seeing}, and in machine learning more broadly~\cite{Chen2026hypo, chen2025auto, Du2026, ke2026view2space}.

\section*{Acknowledgment}

This work was supported by the National Research Foundation, Singapore, under its Competitive Research Programme (NRF CRP24-2020-0003) and by both the National Research Foundation, Singapore, and A*STAR under the Quantum Engineering Programme (NRF 2021-QEP2-02-P12).

\bibliographystyle{IEEEtran}
\bibliography{references}

\end{document}